\documentclass[preprint,nofootinbib, amsmath,amssymb,fontenc, aps]{revtex4-1}

\usepackage{graphicx}
\usepackage{dcolumn}
\usepackage{bm}
\usepackage{hyperref}
\usepackage[mathlines]{lineno}
\usepackage{float}
\usepackage{subcaption}
\usepackage{cleveref}

\newcommand{\beq}{\begin{equation}}
\newcommand{\eeq}{\end{equation}}

\def\lap{\lower.5ex\hbox{$\; \buildrel < \over \sim \;$}}
\def\gap{\lower.5ex\hbox{$\; \buildrel > \over \sim \;$}}

\def\be{\begin{equation}}
\def\ee{\end{equation}}
\def\ba{\begin{eqnarray}}
\def\ea{\end{eqnarray}}

\begin{document}

\title{Clustering of cosmic string loops}

\author{Mudit Jain and  Alexander Vilenkin}

\address{Institute of Cosmology, Department of Physics and Astronomy, Tufts University,
 Medford, MA 02155}

\begin{abstract}

Observational effects of cosmic string loops depend on how loops are distributed in space.  Chernoff \cite{Chernoff} has argued that loops can be gravitationally captured in galaxies and that for sufficiently small values of $G\mu$ their distribution follows that of dark matter, independently of the loop's length. We re-analyze this issue using the spherical model of galaxy formation with full account taken of the gravitational rocket effect -- loop accelerated motion due to asymmetric emission of gravitational waves. We find that only loops greater than a certain size are captured and that the number of captured loops is orders of magnitude smaller than estimated by Chernoff.

\end{abstract}

\maketitle

\section{Introduction}

Cosmic strings are linear topological defects that could be formed at a phase transition in the early universe.  They are predicted in a wide class of particle physics models and can give rise to a variety of observational effects.  Strings can act as gravitational lenses and can produce discontinuous temperature changes and a B-mode polarization pattern on the CMB sky.  Oscillating loops of string emit gravitational waves -- both bursts and a stochastic background.  They can also be sources of synchrotron radiation and of ultrahigh-energy cosmic rays.  
Some superstring-inspired models suggest that fundamental strings may also have astronomical dimensions and play the role of cosmic strings.  String formation, evolution, and observational effects have been extensively studied in the literature (for a review and references see \cite{Book,Tanmay,ChernoffTye}). 

Many observational predictions of cosmic strings depend on how oscillating loops are distributed  in space.  In most of the literature it is assumed that the loop distribution is uniform and is not correlated with galaxies.  The loops formed by the evolving string network initially have relativistic speeds.  They are slowed down by the expansion of the universe, but towards the end of their life they are accelerated due to the generally asymmetric emission of gravitational waves -- the so-called rocket effect.  It is usually assumed that the resulting loop velocities are too high for loops to be captured in cosmic structures.  A notable exception to this view is the work of Chernoff \cite{Chernoff}, who argued that string loops can in fact be captured by galaxies, especially if the strings are sufficiently light (that is, have a small mass per unit length).  In particular, he finds that the number density of loops in our Galaxy could be enhanced by a large factor ($\sim 10^5$) compared to their density in the intergalactic space.  
The loops could then be much closer to the Earth than they would otherwise be, and their observational effects, such as microlensing of stars \cite{ChernoffTye,ChernoffTye2,Chernoff3} or gravitational waves \cite{Hogan,Hogan2}, could be more pronounced.    
 
Since Chernoff's work of 2009, no independent analysis of loop clustering in galaxies has been performed.  In view of its importance for observational predictions, we believe that such an analysis would certainly be useful.  In the present paper we revisit the problem of loop clustering in dark matter halos using the spherical top-hat model of halo formation \cite{GunnGott,Bertschinger}.  This model (which was also used by Chernoff) is not entirely realistic, as it predicts the halo density profile that differs from the Navarro-Frenk-White profile \cite{NFW} suggested by N-body simulations.  The advantage of this model is its simplicity.  Moreover, the coarse halo properties it predicts fit reasonably well with the simulations \cite{Suto}.

Our results are significantly different from those of Chernoff.  In particular, we find that (1) there is a lower bound on the size of loops that get captured in halos, (2) the number of loops that end up in halos is orders of magnitude smaller than Chernoff's estimate, and (3) there are hardly any loops that get captured if their mass parameter is $G\mu \gtrsim 10^{-12}$. The main reason for these discrepancies is the different treatment of the gravitational rocket effect.  Chernoff neglects the role of this effect in loop capture, assuming that it can only be important for ejection of loops from galaxies. Loops are formed with large initial velocities, but then they are slowed down by Hubble expansion, and (neglecting the rocket effect) by the time of galaxy formation they become nearly comoving, so Chernoff finds that the distribution of loops closely follows that of dark matter.  He then shows that the rocket effect fails to eject the captured strings, provided that the strings are sufficiently light.  On the other hand, we find that the rocket effect gives loops significant velocities which depend on the loop's length.  Smaller loops move faster, and loops below a certain size move too fast to be captured in galaxies.  Smaller loops are also more numerous, and a lower cutoff on the loop size implies that only a small fraction within the comoving halo can be captured.

The paper is organized as follows.  In the next section we review the spherical collapse model and discuss the choice of model parameters that we are going to use to represent dark matter halos.  Sec.~\ref{sec:LCorderofmagnitude} begins with a brief review of string evolution and gives a qualitative, order-of-magnitude analysis of loop capture in collapsing halos. Then, a rigorous analytic treatment (which confirms our order of magnitude estimates) of loop capture is provided in Sec.~\ref{sec:LCexactresults}, along with comparisons with our numerical simulation that is laid out in Sec.~\ref{sec:Simulation}. Finally, our conclusions are summarized and discussed in Sec.~\ref{sec:Conclusions}.  In particular, we discuss the differences between our results and those of Chernoff and how our conclusions could be affected by taking into account the hierarchical nature of galaxy formation (we argue that this would not have much of an effect).

\section{Spherical collapse model}
\label{sec:SCmodel}

\subsection{Model outline}

We consider the evolution of a uniform spherical overdensity in a matter-dominated, $\Omega=1$ (Einstein-deSitter) universe.  Following the notation in Ref.~\cite{Bertschinger}, we assume that at some early time $t_i$ the density is 
\beq
\rho=\frac{1}{6\pi Gt_i^2}\equiv \rho_i
\eeq
for $r>R_i$ and $\rho=\rho_i(1+\delta_i)$ with $\delta_i\ll 1$ for $r<R_i$.  We also assume an unperturbed Hubble flow at $t_i$: $v_i=H_i r_i$ with $H_i=2/3t_i$. Also, we only work with leading order terms in $\delta_i$.

The evolution of a comoving spherical shell $r(t)$ of initial radius $r_i$ can be expressed in a parametric form as
\beq
\frac{r}{r_i}\Delta(r_i)=\beta(\theta),
\label{rtheta}
\eeq
\beq
\frac{t}{t_i}=d(\theta)\Delta^{-3/2}(r_i),
\label{ttheta}
\eeq
where 
\beq
\Delta(r_i)=\delta_i \begin{cases}1 & (r_i < R_i)\\ (R_i/r_i)^3 & (r_i > R_i)\end{cases}, 
\label{Delta}
\eeq
$\beta(\theta)=\sin^2(\theta/2)$, and $d(\theta)=(3/4)(\theta-\sin\theta)$.

The shell reaches the maximum (turnaround) radius $r_{ta}$ at $\theta=\pi$ and begins to collapse.  The turnaround radius and time for a given shell can be found from
\beq
r_{ta}=\begin{cases}\frac{r_i}{\delta_i} & (r_i < R_i)\\ \frac{r_i^4}{\delta_i R_i^3} = R_{ta}\left(\frac{t_{ta}}{T_{ta}}\right)^{8/9} & (r_i > R_i)\end{cases},
\label{rv}
\eeq
where $R_{ta}=R_i/\delta_i$ and $T_{ta}=d(\pi) \delta_i^{-3/2} t_i$ are the turnaround radius and time of the initial overdense shell.  We shall assume that a collapsing shell virializes and stops evolving when it contracts to $r_v=r_{ta}/2$, which corresponds to $\theta=3\pi/2$. The mass profile $M(r)$ at $0<r<r_v(t)$ is now fixed and is given by
\beq
M(r)=\frac{4\pi}{3}\rho_i r_i^3=M_0\begin{cases}\left(\frac{r}{R_v}\right)^{3/4} & (R_v<r<r_v(t))\\
\left(\frac{r}{R_v}\right)^3 & (0<r<R_v)\end{cases},
\label{Mr}
\eeq
where $M_0 \approx (4\pi/3)\rho_i R_i^3$ is the mass within the initial overdensity. We can also express the virialized mass as a function of redshift at $z<z_v$:
\beq
M_v(z)=M_0 \left(\frac{r_v(t)}{R_v}\right)^{3/4} = M_0\left(\frac{t}{T_v}\right)^{2/3} = M_0\left(\frac{1+z_v}{1+z}\right).
\label{Mvz}
\eeq 
(This expression is not expected to be valid at $z\lesssim 1$, when the cosmological constant begins to dominate.)
As we have mentioned, the mass profile (\ref{Mr}) is different from the NFW profile suggested by N-body simulations, which gives $M(r)\propto \ln r$ at large $r$.  We will see, however, that the region far outside of the top hat does not play much of a role in loop capture. 

The mass of a top hat halo virializing at time $T_v$ with a radius $R_v$ can be expressed as
\beq
M_0 = \frac{4\pi}{3}\rho_i R_i^3 = \frac{16}{9} D^2 \frac{R_v^3}{GT_v^2},
\label{M0}
\eeq
where 
\beq
D\equiv d(3\pi/2)\approx 4.3.
\label{D}
\eeq
It will also be convenient to express $M_0$ in terms of the turnaround parameters:
\beq
M_0=\frac{\pi^2}{8}\frac{R_{ta}^3}{GT_{ta}^2}.
\label{M0ta}
\eeq
The top hat halo density at the time of turnaround is
\beq
\rho_{tophat} = \frac{3M_0}{4\pi R_{ta}^3} = \frac{3\pi}{32GT_{ta}^2},
\eeq
and its overdensity compared to the FRW background at $t=T_{ta}$ is
\beq
\frac{\rho_{tophat}}{\rho_{FRW}}=\frac{9\pi^2}{16} \approx 5.5.
\label{overdensity}
\eeq
We finally give the following useful relation between the turnaround and virialization times and the corresponding redshifts:
\beq
\frac{T_{ta}}{T_v}=\left(\frac{1+z_v}{1+z_{ta}}\right)^{3/2} =\frac{3\pi}{4D}=0.55.
\label{tav}
\eeq

\subsection{Choice of parameters}

During the epoch of interest to us here, our universe is accurately described by the LCDM model, while the spherical collapse model of the preceding subsection assumes a flat matter-dominated (Einstein-de Sitter) universe.  On the other hand, halos that we are interested in, collapse at $z\gtrsim 2$ when the Einstein-de Sitter model gives a reasonably accurate approximation. To make a connection between the two models, we use the LCDM scale factor
\beq
a(t)=\left(\frac{\Omega_m}{\Omega_{vac}}\right)^{1/3} \sinh^{2/3} \left(\frac{3}{2} H_{vac} t\right) = (1+z)^{-1},
\label{at}
\eeq
where $\Omega_m\approx 0.3$ is the present matter density parameter, $\Omega_{vac}=1-\Omega_m$, $H_{vac}=\sqrt{\Omega_{vac}} H_0 \approx 0.84 H_0$, and $H_0\approx 67~{\rm km/s\cdot Mpc}$ is the present Hubble parameter. For $H_{vac} t\ll 1$ this gives
\beq
a(t)\approx (9\Omega_m/4)^{1/3} (H_0 t)^{2/3}
\label{atapprox}
\eeq
and
\beq
\frac{3}{2} \sqrt{\Omega_m} H_0 t \approx (1+z)^{-3/2}.
\eeq
With $\Omega_m\approx 0.3$ we have
\beq
H_0 t\approx 1.2 (1+z)^{-3/2}.
\label{connection}
\eeq

To assess the validity of the approximations (\ref{atapprox}) and (\ref{connection}), we note that keeping only the first term in the expansion
\beq
\sinh x = x + \frac{x^3}{6} + ...
\eeq
is accurate within $\sim x^2 /6$.  On the other hand, for $z\sim 2$ we have $3H_{vac} t/2 \sim 0.25$, so (\ref{atapprox}) is accurate within $\sim 1\%$.  The accuracy is even better at higher redshifts.

Our halo formation model is specified by two parameters: $R_v$ and $z_v$.  We set $R_v=60~$kpc and $z_v=3$ as representative values.  With this choice, $M_0 \approx 3.2\times 10^{11} M_\odot$ and the virialized mass $M_v(z)$ in Eq.~(\ref{Mvz}) is roughly consistent with the mass assembly history for the Milky Way at $z\lesssim 3$ \cite{Evans}.  
We note that typical halos virializing at $z=3$ have significantly smaller masses, $\sim 10^9 M_\odot$.  
Using the observationally suggested power spectrum of density fluctuations, as given in \cite{Komatsu},
it can be shown that our value of $M_0$ corresponds to $\sim 2\sigma$ mass fluctuation in the top hat. It should be noted that our choice of parameters is somewhat imprecise, since the mass assembly history and the density profile predicted by the spherical model are not accurate fits to observations or to N-body simulations.\footnote{Chernoff {\it et. al.} pointed out that in the relevant range of radii the density profile of the Milky Way can roughly be fitted by a power law $\rho(r) \approx 10^9 r^{-9/4} M_\odot /{\rm kpc}^3$.  This agrees with the profile predicted by the spherical model, $\rho(r)\approx 1.4\times 10^3 (1+z_v)^3 (R_v/r)^{9/4} M_\odot /{\rm kpc}^3$, for our choice of parameters.} On the other hand, the spherical model has been successfully used to account for many aspects of nonlinear dynamics of structure formation, so one can expect that it should work reasonably well for an {approximate} analysis of loop capture.  We will further comment on this in Sec.~\ref{sec:Conclusions}.

\section{Loop capture I: order of magnitude estimates}
\label{sec:LCorderofmagnitude}

\subsection{String evolution}

Numerical simulations of string evolution indicate that strings evolve in a self-similar manner. A Hubble-size volume at any time $t$ contains a few long strings stretching across the volume and a large number of closed loops of length $l \ll t$ (for an up to date review of string simulations, see \cite{BOS16})\footnote{Here $l$ is the so-called invariant length of the loop, defined as $l=E/\mu$, where $E$ is the loop's center of mass energy}. 
Long strings move, typically at mildly relativistic speeds ($v\sim 0.2$) and reconnect when they cross.  Reconnections lead to the formation of closed loops.  The loops oscillate periodically and emit gravitational radiation at the rate
\beq
{\dot E}= \Gamma G\mu^2,
\eeq
where $G$ is Newton's constant, $\mu$ is the mass per unit length of string, and $\Gamma\sim 50$ is a numerical factor depending on a particular loop configuration.  As loops loose their energy, they gradually shrink and eventually disappear.  The lifetime of a loop of initial length $l$ is
\beq
\tau=\frac{l}{\Gamma G\mu}.
\eeq
$G\mu$ is an important dimensionless parameter characterizing the strength of gravitational interaction of strings. 
Gravitational waves emitted by loops over the cosmic history add up to a stochastic gravitational wave background.   Requiring that the predicted amplitude of this background is not in conflict with the millisecond pulsar observations, one can impose an upper bound on the string parameter $G\mu$ \cite{BOS18}:
\beq
G\mu\lesssim 10^{-11}.
\eeq

Loops of interest to us were formed in the radiation era.  A loop formed at time $t_f$ has length $l\sim 0.1 t_f$.\footnote{Much smaller loops are also produced in localized regions where the long string velocity approaches the speed of light.  Such loops decay soon after they are formed and will be of no interest to us here.} The smallest loops surviving at the present time $t_0$ have lifetime $\tau\sim t_0$ and initial length
\beq
{l_*}\sim \Gamma G\mu t_0.
\label{l*}
\eeq
They were formed at $t_f\sim 10 {l_*}$.  It will be convenient to characterize the loop length by a dimensionless number $\xi=l/{l_*}$.  Then the loop formation time is
\beq
t_f\sim 10\xi \Gamma G\mu t_0.
\eeq
The average number density for large loops ($\xi\gg 1$) of size $\sim l$ at redshift $z$ in the matter era is\footnote{Note that here we use the definition $n(\xi)=\xi (dn/d\xi)$, which is the loop density per logarithmic interval of length.  This is different from \cite{BO17}, where the notation $n(l)$ is used for what we denote $dn/dl$.} \cite{BO17}
\beq
n(z,\xi)\sim 0.5 \frac{(H_0^2 \Omega_{r0})^{3/4} (1+z)^3}{l^{3/2}} \sim 10^{-6} (G\mu)^{-3/2} \xi^{-3/2} t_0^{-3} (1+z)^3,
\label{nzxi}
\eeq
where $\Omega_{r0}=9\times 10^{-5}$ is the density fraction in massless (light) particles, including neutrinos. Loops are chopped off the long string network with initial velocity $v_f\sim 0.3$, which, relative to the background Hubble flow, gets reduced and becomes
\beq
{\bf v}_0(t)\sim {\bf v}_f (t_f/t_{eq})^{1/2} (t_{eq}/t)^{2/3} = 2.6\,\left(1+z\right)\left({\xi G\mu}\right)^{1/2}{\bf v}_f
\label{v0}
\eeq
in the matter era. Here, $t_{eq} \approx 2H_0^{-1}\Omega_m^{-1/2}(1+z_{eq})^{-3/2}$ is the time of equal radiation and matter densities\footnote{We have verified that with this definition of $t_{eq}$ Eq.~(\ref{v0}) gives an accurate transition from radiation to matter eras.} and $z_{eq}\approx 3440$ is the corresponding redshift.

The loop motion is also affected by the rocket effect \cite{HoganRees,VV}.  Emission of gravitational waves by a loop is generally asymmetric, resulting in a recoil force on the loop $F\sim \Gamma_p G\mu^2$, where $\Gamma_p \sim 0.1 \Gamma$ \cite{VV}.
Hence the loop equation of motion is
\begin{equation}
\dot{\bf v}_{pec} + H{\bf v}_{pec} =\dfrac{\Gamma_p}{\Gamma\,\xi\,t_0}{\bf n}
\label{loopeom1}
\end{equation}
where ${\bf n}$ is the unit vector in the direction of the rocket force, and ${\bf v}_{pec}$ is the loop's peculiar velocity. The solution of Eq.~(\ref{loopeom1}) is
\beq
{\bf v}_{pec}(t)={\bf v}_0(t) +\frac{3}{5} \frac{\Gamma_p}{\Gamma}\frac{t}{\xi t_0}{\bf n}
\label{vsol}
\eeq
with ${\bf v}_0(t)$ from Eq.~(\ref{v0}). \footnote{Here we have assumed FRW cosmology, i.e. $H = 2/3t$. In the next section we shall improve upon this and analyze loop dynamics within the top hat rigorously.}

The first term in Eq.~(\ref{vsol}) decreases with time, while the second (rocket) term grows with time.  The two terms become comparable at time
\beq
t_r\sim 5\xi^{9/10} (G\mu)^{3/10} t_0
\eeq
or redshift
\beq
1+z_r \sim 0.4 \xi^{-3/5} (G\mu)^{-1/5},
\label{zr}
\eeq
and the rocket term dominates afterwards. For small values of $G\mu$ and $\xi$ not very large,\footnote{We are interested in the smallest relevant values of $\xi$, since the loop density (\ref{nzxi}) decreases with $\xi$.} this happens at $z_r \gg z_v$, where $z_v$ is the redshift of halo virialization.  Then we can disregard the first term in Eq.~(\ref{vsol}) for the loop velocity and use
\beq
{\bf v}_{pec}(t) \sim 0.06  \frac{t}{\xi t_0}{\bf n} .
\label{vt}
\eeq
This approximation applies for
\beq
\xi \ll \xi_r \sim 200 \mu_{-12}^{-1/3}\left(\frac{1+z_v}{4}\right)^{-5/3},
\label{xi<xir}
\eeq
where $\mu_{-12}\equiv G\mu/10^{-12}$. We shall verify that for observationally allowed values of $G\mu$ {almost all} of the captured loops satisfy this condition (see discussion in Sec.~\ref{sec:Conclusions}).

\subsection{Loop capture within the top hat}

We shall first consider loop capture in the top hat halo.  We need to compare the loop velocity $v_{pec}$ to the escape velocity from the halo, $v_{esc}\sim (2GM_0/R)^{1/2}$, where $R$ is the top hat radius.  Both $v_{pec}$ and $v_{esc}$ are time-dependent: the rocket velocity grows with time, while the escape velocity decreases as the halo expands.  We shall therefore impose the capture condition, $v_{pec} < v_{esc}$ at the turnaround time $T_{ta}$.

For a rough estimate, we shall assume that the loop velocity at $t\lesssim T_{ta}$ is not much affected by the halo evolution and is given by Eq.~(\ref{vt}) with $t\sim T_{ta}$.  Then, using Eq.~(\ref{connection}), we have
\beq
v_{pec}(z_{ta})\sim 0.06 \frac{T_{ta}}{\xi t_0}.
\eeq
The escape velocity from the halo is
\beq
v_{esc}\sim \left(\frac{2GM_0}{R_{ta}}\right)^{1/2}=\frac{\pi}{2}\frac{R_{ta}}{T_{ta}}.
\label{vesc}
\eeq
Requiring that $v_{pec}<v_{esc}$, we obtain a lower bound on the size of captured loops:
\beq
\xi \gtrsim \xi_{min} \sim 0.04 \frac{T_{ta}^2}{R_{ta}t_0}\sim 9.2 \left(\frac{R_v}{60{\rm kpc}}\right)^{-1} \left(\frac{1+z_v}{4}\right)^{-3}.
\label{xibound}
\eeq
where we have used Eqs.~(\ref{connection}) and (\ref{tav}). A more accurate estimate of $\xi_{min}$ will be given in Sec.~\ref{subsec:dynamics_tophat}, with the numerical coefficients 0.04 and 9.2 in (\ref{xibound}) replaced by $\sim 0.1$ and $\sim 25$ respectively. We will use these improved values in the rest of this section.

It follows from the second step in Eq.~(\ref{vesc}) that loops with $v_{pec}<v_{esc}$ do not have enough time to cross the halo at $t\sim T_{ta}$.  Since $v_{pec}\propto t$ and the halo size at $t\ll T_{ta}$ is $R\propto t^{2/3}$, the ratio $v_{pec}t/R$ is even smaller at earlier times.  This indicates that the loops that get captured (that is, having $\xi>\xi_{min}$) are essentially comoving: their number within the top hat remains approximately constant until the turnaround.

The number of loops captured within the turnaround radius can be estimated simply as their number within the top hat halo at $z=z_i$:
\beq
N_{ta}(\xi)\sim \frac{4\pi}{3} n(z_i,\xi) R_i^3 \sim 10^{-16} (G\mu\xi)^{-3/2} \left(\frac{R_v}{60{\rm kpc}}\right)^3 \left(\frac{1+z_v}{4}\right)^3.
\eeq
For $z_v=3$ and $R_v=60$~kpc we find
\beq
N_{ta}(\xi)\sim 0.8 \mu_{-12}^{-3/2} \left(\frac{\xi_{min}}{\xi}\right)^{3/2} ,
\label{N3}
\eeq
where we have used $\xi_{min}\sim 25$. Combined with $\xi\gtrsim \xi_{min}$, this indicates that a substantial number of loops ($\gtrsim 10^3$) get captured in top hat halos for $G\mu\lesssim 10^{-14}$, while we do not expect any loops to be captured for $G\mu\gtrsim 10^{-12}$. Most of the captured loops are expected to have the smallest size, $\xi\sim\xi_{min}$.  

At the time of turnaround, dark matter particles have zero velocity and later collapse to virialize at radius $\sim R_v = R_{ta}/2$.  But captured loops have velocities up to $v_{esc}$ and we expect them to settle into orbits of radii up to $\sim R_{ta}$.

\subsection{Capture outside of top hat}

Let us now consider a loop which is initially at a radius $r_i >R_i$ outside of the top hat.  The shell of initial radius $r_i$ turns around at time
\beq
t=T_{ta}\left(\frac{r_i}{R_i}\right)^{9/2}.
\eeq
Its turnaround radius is
\beq
r_{ta}=R_{ta}\left(\frac{r_i}{R_i}\right)^4= R_{ta}\left(\frac{t}{T_{ta}}\right)^{8/9}.
\label{rtat}
\eeq
The mass enclosed by the shell is
\beq
M(t)=M_0 (t/T_{ta})^{2/3}
\label{Mt}
\eeq
and the escape velocity from its outer region is
\beq
v_{esc}(t)= \left(\frac{2GM(t)}{r_{ta}(t)}\right)^{1/2} = v_{esc}^{(0)} (T_{ta}/t)^{1/9},
\eeq
where $v_{esc}^{(0)}$ is the escape velocity from the top hat, given by Eq.~(\ref{vesc}).  

We note that the combination $v_{esc}(t)\cdot t/r_{ta}(t)\sim 1$ is independent of time.  This implies that loops with $v<v_{esc}$ in the outer region $r\sim r_{ta}$ do not have time to cross that region at turnaround.  By the same argument as before, such loops are nearly comoving up to the turnaround time.
Requiring that $v_{pec}(t)<v_{esc}(t)$ at turnaround, we obtain the capture condition 
\beq
\xi> \xi_{min} (t/T_{ta})^{10/9} = \xi_{min} (r_{ta}/R_{ta})^{5/4},
\label{eq:ximinoutside}
\eeq
where $\xi_{min}$ is the minimal captured loop size for the top hat, given by Eq.(\ref{xibound}).  The initial radius of the region within which all loops of a given size $\xi$ are captured is then given by 
\begin{equation}
r_i \sim R_i\left(\dfrac{\xi}{\xi_{min}}\right)^{1/5}
\end{equation}
and the total number of captured loops of size $\xi$ is
\begin{eqnarray}
N(\xi) \sim \dfrac{4\pi}{3}r_i^3\,n(z_i,\xi)
&\approx& 1.0\,\mu_{-12}^{-3/2}\left(\dfrac{R_v}{60\,\text{kpc}}\right)^{9/2}\left(\dfrac{1+z_v}{4}\right)^{15/2}\left(\dfrac{\xi_{min}}{\xi}\right)^{9/10}.
\label{Ntot}
\end{eqnarray}
This decreases with $\xi$ slower than the number of loops captured within the top hat halo (\ref{N3}), so most of the loops with $\xi>\xi_{min}$ are to be found outside of the halo.

We shall assume that loops turning around at $r\sim r_{ta}$ end up in orbits of radii $r \sim r_{ta}$.  Then the loops at a distance $r$ from the center of the halo have typical size $\xi(r)\sim \xi_{min}(r/R_{ta})^{5/4}$.
The density of such loops is 
\beq
n(r)\sim n\left(z_i,\xi(r)\right)r_i^3/r^3 \propto r^{-33/8},
\label{nr}
\eeq
where we have used $n(z,\xi)\propto \xi^{-3/2}$, $\xi(r) \propto r^{5/4}$, and $r_i \propto r^{1/4}$. The loop density in Eq.~(\ref{nr}) decreases faster than $r^{-3}$, so the total number of loops $N(r)$ within radius $r$ does not significantly increase with the radius.  

In a realistic LCDM cosmology, structure formation effectively ceases when the cosmological constant $\Lambda$ starts dominating at a redshift $z_{\Lambda} \sim 0.3$. The last shell that turns around at this time has initial radius
\begin{equation}
r_{i\Lambda} \sim \left(\dfrac{1+z_v}{1 + z_{\Lambda}}\right)^{1/3}R_i,
\label{ri_lambda}
\end{equation}
within which loops of the following size are captured (cf. \eqref{eq:ximinoutside})
\begin{equation}
\xi_{\Lambda} \sim \xi_{min}\left(\dfrac{1+z_v}{1+z_{\Lambda}}\right)^{5/3}.
\end{equation}
Hence we expect Eq.~(\ref{Ntot}) to apply for $\xi_{min}\lesssim \xi \lesssim \xi_\Lambda$.  For $\xi>\xi_{\Lambda}$, all captured loops are contained within the same initial radius $r_{i\Lambda}$ and their number is 
\beq
N(\xi>\xi_\Lambda)\propto \xi^{-3/2}.
\label{NtotLambda}
\eeq
The estimates for $N_{ta}(\xi)$ and $N(\xi)$ that we obtained in this section are in a good agreement with our numerical simulations (see Sec.~\ref{sec:Simulation}).

\subsection{Loop ejection}

The captured loops can be ejected from the galaxy due to the rocket effect.  This happens if the rocket acceleration gets larger than the gravitational acceleration.  For a loop orbiting the halo at radius $\sim r$, this condition is
\beq
\frac{0.1}{\xi t_0}\gtrsim \frac{GM(r)}{r^2}= \frac{GM_0}{R_v^2}\left(\frac{R_v}{r}\right)^{5/4},
\eeq
where in the last step we used Eq.~(\ref{Mr}) with $r>R_v$.  This implies that loops at $r>r_{max}(\xi)$ will be ejected, where the maximal radius $r_{max}(\xi)$ is given by
\beq
r_{max}(\xi)\sim R_v\left(\frac{10GM_0 t_0\xi}{R_v^2}\right)^{4/5}\sim 3.6 R_v\left(\frac{\xi}{\xi_{min}}\right)^{4/5}.
\label{rtr}
\eeq
Since all captured loops satisfy $\xi>\xi_{min}$, the maximal radius is $r_{max}\gtrsim 3.6 R_v$.

As the loops evaporate, the length parameter $\xi$ decreases, the rocket acceleration increases, and all loops eventually get ejected.  The characteristic timescale for this process however, is rather long:
\beq
t_{esc}\sim \frac{\xi}{\dot\xi} \sim \xi t_0.
\eeq
Since $\xi\gtrsim 25$, we do not expect significant loop ejection by the present cosmic time.  

\section{Loop capture II: exact results}
\label{sec:LCexactresults}

\subsection{Loop potential energy}
\label{subsec:LPE}

The potential felt by a loop of size $\xi$ is
\begin{equation}
V(s,\alpha,t) = G\int^{s}_{0}dr\,\dfrac{M(r,t)}{r^2} - \dfrac{H_0}{\xi}\dfrac{\Gamma_p}{\Gamma}\,s\,\cos\alpha
\label{potential}
\end{equation}
where $s$ and $\alpha$ are respectively the radial distance of the loop and the angle between the rocket direction and the radius vector from the center of the top hat.  As before, $M(r,t)$ is the mass enclosed within the radius $r$.
The last term in Eq.~(\ref{potential}) accounts for the force due to the rocket.  

The virialization radius at time $t>T_v$, where $T_v$ is the virialization time of the top hat, is given by (see \eqref{rv})
\begin{equation}
r_{v}(t) = R_v\left(\dfrac{t}{T_v}\right)^{8/9}.
\label{eq:energy.conserving.shell}
\end{equation}
For $r<r_v(t)$ the background density has virialized and therefore the loop potential is time-independent. The mass function is then given by Eq.~(\ref{Mr}) and the potential (\ref{potential}) becomes
\begin{eqnarray}
V(s,\alpha) &=& \dfrac{GM_0}{2R_v^3}\left[s^2 \Theta\left(R_v - s\right) + \left(9R_v^2 - 8\dfrac{R_v^{9/4}}{s^{1/4}}\right)\Theta\left(s - R_v\right)\right]. \nonumber\\
&& - \dfrac{H_0}{\xi}\dfrac{\Gamma_p}{\Gamma}\,s\,\cos\alpha\nonumber\\
\label{Vsalpha}
\end{eqnarray}

\begin{figure}[t]
\centering
\includegraphics[width=0.75\columnwidth]{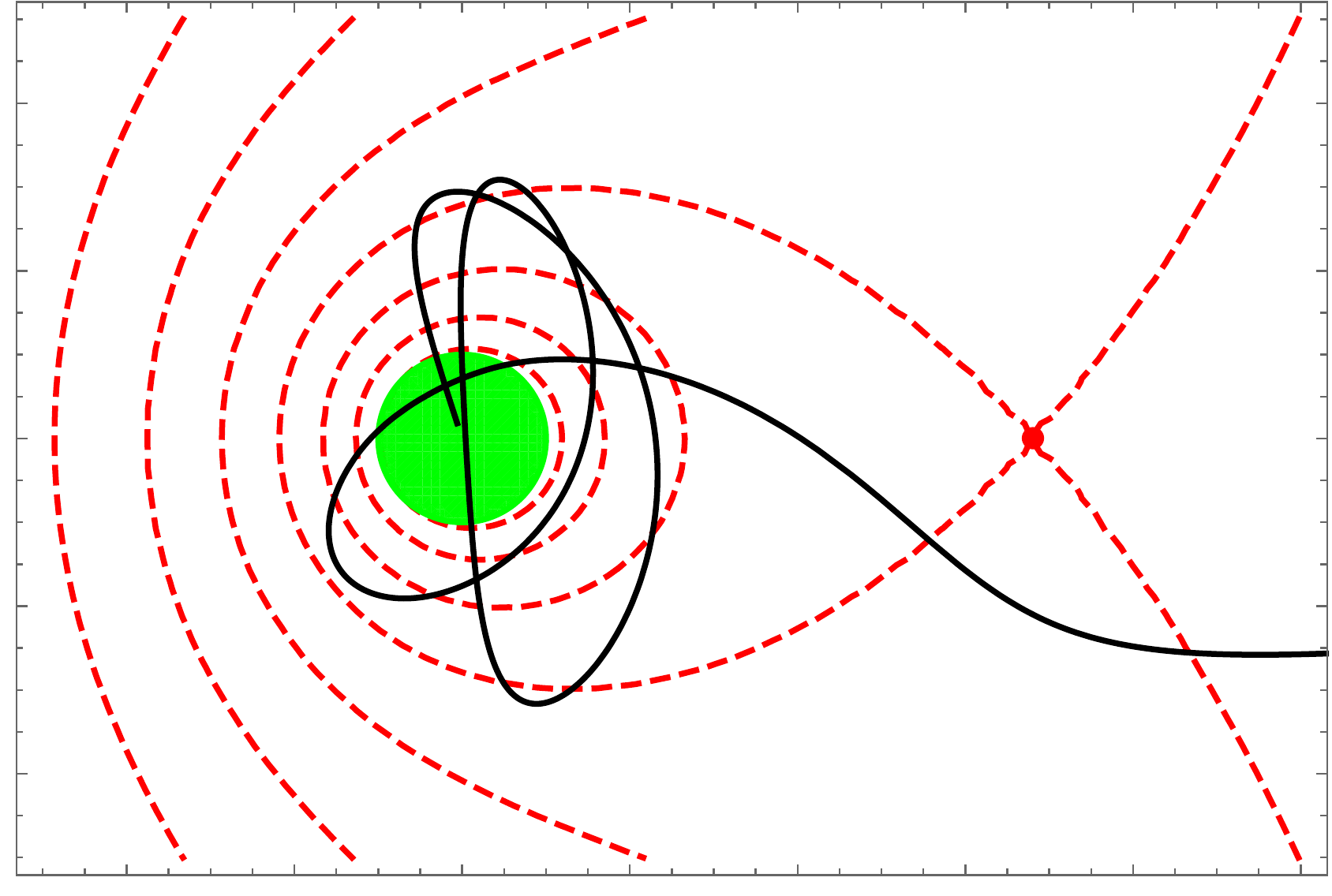}\vspace{0.1cm}\\
\caption{A plot showing equipotential surfaces of the virialized potential \eqref{Vsalpha} for $\xi = 11\,\xi_*$ (shown in dashed red) with the rocket pointing to the right along the horizontal axis. The virialized top hat halo is shown in green and the saddle point of the potential is marked by a red dot.  The black curve is the trajectory, as obtained from simulation (see Sec.~\ref{sec:Simulation}), for a loop which is temporarily captured by a collapsing halo but eventually escapes.}
\label{fig:contour.plot}
\end{figure}

For a loop to be captured, the potential must have a local minimum.  {It is clear from Eq.~(\ref{Vsalpha}) that the minimum can only occur at $\alpha=0$, and one can easily verify that it exists} only if $\xi$ obeys a lower bound:
\begin{equation}
\xi > \left(H_0\,R_v\right)\left(\dfrac{R_v}{G M_0}\right)\left(\dfrac{\Gamma_p}{\Gamma}\right) \equiv \xi_{*}.
\end{equation}
In this case the potential has a saddle point at $\alpha=0$ and $s=s_{max}$ with
\begin{equation}
s_{max} = \left(\dfrac{\xi}{\xi_{*}}\right)^{4/5}R_v.
\label{eq:rmax}
\end{equation}
The pattern of equipotential surfaces in the vicinity of the top hat is illustrated in Fig.~\ref{fig:contour.plot} for $\xi>\xi_*$.
The potential value at the saddle point is
\begin{equation}
E_{B} = V(s_{max},0) = \dfrac{GM_0}{2R_v}\left(9 - 10\left(\dfrac{\xi_{*}}{\xi}\right)^{1/5}\right),
\label{bounding.energy}
\end{equation}
which is the maximal energy that a captured loop can have.  The corresponding equipotential surface bounds the region where captured loops with a given rocket direction can be located. Note that $s_{max}$ in Eq.~(\ref{eq:rmax}) is basically the same as $r_{max}$ in Eq.~(\ref{rtr}) -- which is not surprising: in both cases it is the maximal radial distance that a captured loop can have from the center of the halo.

We also note that some loops can be temporarily captured even if they have energy $E>E_B$. As an example in Fig.~\ref{fig:contour.plot}, we show the trajectory of a loop of size $\xi = 11\xi_*$ which orbits a few times about the halo before eventually escaping.  Our numerical simulations (described in Sec.~\ref{sec:Simulation}) indicate that this behavior is rare and requires a rather fine-tuned initial position vector of the loop.  Most of the loops are either captured or escape without first orbiting the halo. 

\subsection{Loop dynamics within the top hat}
\label{subsec:dynamics_tophat}

In this section, we re-analyze the dynamics of loops which stay within the top hat until it virializes, rigorously. Observationally, this is an interesting set of loops since our Solar system is well within the virialization radius of the Galaxy, and most significant observational effects are expected to come from nearby loops.  
All loops that stay within the top hat until $t = T_v$ obey the previous Eq.~\eqref{loopeom1}, but with the Hubble parameter given by
\begin{equation}
H = \dfrac{2\,\delta^{3/2}}{3\,t_i}\dfrac{\cot(\theta/2)}{\beta(\theta)}
\end{equation}
in terms of $\theta$ (cf. \eqref{rtheta} and \eqref{ttheta}). Using Eqs.~(\ref{ttheta}) and (\ref{connection}), Eq.~(\ref{loopeom1}) can be recast as
\begin{equation}
\dfrac{d{\bf v}_{pec}}{d\theta} + {\bf v}_{pec}\cot\left(\theta/2\right) = \dfrac{0.18}{d\left(3\pi/2\right)}\dfrac{\beta(\theta)}{\xi(1+z_v)^{3/2}}{\bf n},
\label{loopeom3}
\end{equation}
the solution to which is 
\begin{equation}
{\bf v}_{pec}\left(\theta\right) = \dfrac{2\,R_v\delta^{3/2}}{3\,t_i}\left(\dfrac{\xi_*}{\xi}\right)\left(\dfrac{6\theta - 8\sin\theta + \sin 2\theta}{4\,\beta(\theta)}\right){\bf n} + \left(\dfrac{\delta}{\beta(\theta)}\right)\,{\bf v}_{0}(t_i).
\label{eq:peculiar.velocity}
\end{equation}
Here ${\bf v}_{0}(t_i)$ is the initial peculiar velocity of the loop (early on in the matter era) given by Eq.~\eqref{v0}.

We note in passing that since we now have precise dynamics of loops in the top hat model, we can provide a more reliable estimate for $\xi_{r}$ (cf. Eq. \eqref{xi<xir}) by comparing the two terms in Eq.~(\ref{eq:peculiar.velocity}) at virialization ($\theta = 3\pi/2$):
\begin{equation}
\xi_{r} \simeq 132\,\mu_{-12}^{-1/3}\left(\dfrac{1+z_v}{4}\right)^{-5/3}.
\label{eq:xi_r}
\end{equation}
This agrees with our estimate \eqref{xi<xir} within a factor of 2. 

The velocity of the loop relative to the top hat center is
\begin{eqnarray}
{\bf v}\left(\theta\right) = {\bf v}_{pec} (\theta) + {\bf v}_H (\theta),
\label{eq:velocity.physical}
\end{eqnarray}
where
\begin{eqnarray}
{\bf v}_H (\theta) = \dfrac{2\,\delta^{3/2}}{3\,t_i}\dfrac{\cot(\theta/2)}{\beta(\theta)}\,{\bf s}(\theta)
\end{eqnarray}
is the Hubble velocity and ${\bf s}(\theta)$ is the position vector of the loop from the origin:
\begin{eqnarray}
{\bf s}(\theta) &=& \dfrac{{\bf s}_{i}}{\delta}\beta(\theta) + \beta(\theta)\int^{\theta}_{\theta_i} d\theta\,\dfrac{dt}{d\theta}\,\dfrac{1}{\beta(\theta)}{\bf v}_{pec}(\theta)\nonumber\\
&=& \dfrac{{\bf s}_{i}}{\delta}\beta(\theta) + R_v\left(\dfrac{\xi_{*}}{\xi}\right)\beta(\theta)\left(\cos\theta - 3\theta\cot(\theta/2) + 5\right)\,{\bf n}.
\label{eq:position.vector}
\end{eqnarray}
Here we have neglected the initial velocity ${\bf v}_{0}(t_i)$, assuming that condition (\ref{eq:xi_r}) is satisfied. 

At the time of top hat virialization $(\theta=3\pi/2)$ we have
\beq
\frac{{\bf s}_v}{R_v}=\frac{{\bf s}_i}{R_i} +\left(\frac{9\pi+10}{4}\right)\frac{\xi_*}{\xi} {\bf n}.
\label{spheres}
\eeq
This equation has a simple geometric interpretation.  Loops that were initially uniformly distributed within a sphere of radius $R_i$ are distributed at $t=T_v$ in a sphere of radius $R_v$ which is displaced from the top hat sphere by {the} vector ${\bf b}=\left(\frac{9\pi+10}{4}\right)\frac{\xi_*}{\xi} R_v {\bf n}$.  In order to have any loops remaining within the top hat at $t=T_v$, we must have $b < 2R_v$.  This yields a lower bound on $\xi$\footnote{{This is the improved estimation of $\xi_{min}$ we mentioned earlier}}:
\begin{eqnarray}
\xi > \dfrac{9\pi+10}{8}\,\xi_* \equiv \xi_{min} \simeq 24.6 \left(\dfrac{R_v}{60\,\text{kpc}}\right)^{-1}\left(\dfrac{1+z_v}{4}\right)^{-3}.
\label{ximin}
\end{eqnarray}
Now let us consider all those loops that stay within the top hat until virialization. The velocities and potential values of such loops at $T_v$ are respectively
\begin{eqnarray}
{\bf v}_{v} &=& -\left(\dfrac{GM_0}{R_v}\right)^{1/2}\left[\left(\dfrac{4}{9\pi+10}\right)\dfrac{\xi_{min}}{\xi}\,{\bf n} + \dfrac{{\bf s}_i}{R_i}\right],\\
\label{vv}
W_{v} &=& \dfrac{GM_0}{2R_v}\left[\left(\dfrac{s_{i}}{R_i}\right)^2 + \left(\dfrac{36\pi+8}{9\pi+10}\right)\left(\dfrac{\xi_{min}}{\xi}\right)^2 + \left(\dfrac{36\pi+24}{9\pi+10}\right)\dfrac{\xi_{min}}{\xi}\,\dfrac{s_{i}}{R_i}\,\cos\alpha_i\right].
\label{eq:position.velocity.virialization.loops.within}
\end{eqnarray}
In order for such loops to be captured, the total energy $E_{v}$ carried by them at $T_v$ must be smaller than the bounding energy $E_{B}$ (cf. \eqref{bounding.energy}):
\begin{eqnarray}
\tilde{E}_v &=& \dfrac{1}{2}v_{v}^2 + W_{v} \approx \dfrac{GM_0}{2R_v}\left[2\left(\dfrac{s_{i}}{R_i}\right)^2 + 3.17\left(\dfrac{\xi_{min}}{\xi}\right)^2 + 3.8\left(\dfrac{\xi_{min}}{\xi}\right)\left(\dfrac{s_{i}}{R_i}\right)\cos\alpha_i\right]\nonumber\\
&\lesssim& \dfrac{GM_0}{2R_v}\left(9 - 7.31\left(\dfrac{\xi_{min}}{\xi}\right)^{1/5}\right)
\label{eq:energy.in}
\end{eqnarray}
This is indeed satisfied for all loops that stay within the top hat until $T_v$. 

It should be noted that this comparison of energies to verify capturing is only strictly valid when the relevant part of the background region has virialized -- that is, if the radius $r_v(t)$ has extended beyond the saddle point of the potential. This is not true at $T_v$, but we note that most of the loops at $T_v$ have their velocities directed towards the center of the halo (see Eq.~(\ref{vv}) with $\xi>\xi_{min}$). By the time these loops cross the halo and emerge on the other side, the virialization radius would extend further out.\footnote{The virialization radius will also extend while some of the loops are bouncing around in the halo region, as discussed at the end of Sec.~\ref{subsec:LPE}.} 
Our numerical simulations, discussed in the next section, indicate that all such loops are indeed captured.

All loops with $\xi < \xi_{min}$ will have necessarily crossed the top hat before $T_v$.  Numerical simulations suggest that all such loops escape to infinity.  There are also some loops with $\xi>\xi_{min}$ which cross out of the top hat before $T_v$.  Some of these loops get captured and some escape.	
Note also that for $\xi\gg\xi_{min}$ the two spheres discussed below Eq.~(\ref{spheres}) nearly overlap, implying that all loops with $\xi\gg\xi_{min}$ that were initially within the top hat will get captured.

\subsection{Number of captured loops within top hat}

With the aid of Eq.~(\ref{spheres}), the condition for loops to remain within the top hat until virialization can be expressed as
\begin{equation}
\left(\dfrac{s_{i}}{R_i}\right)^2 + 4\left(\dfrac{\xi_{min}}{\xi}\right)^2 + 4\cos\alpha_i\left(\dfrac{s_{i}}{R_i}\right)\left(\dfrac{\xi_{min}}{\xi}\right) < 1.
\label{eq:constraint.in.out.2}
\end{equation}
As we have discussed, all such loops are captured, and we expect them to settle into orbits of radii $\lesssim R_v$. The number of such loops having size $\sim\xi$ is equal to
\begin{equation}
N_{\text{tophat}}(\xi) = 2\pi\,n(z_i,\xi) R_i^3\int^{1}_{-1}d\zeta\int^{1}_{0}d\lambda\,\lambda^2\,\Theta\left(\text{constraint}\,\eqref{eq:constraint.in.out.2}\right),
\end{equation}
where again, 
\begin{equation}
n\left(z_i,\xi\right) = \dfrac{H_0^3(1+z_i)^3}{2\,\xi^{3/2}_{min}}\left(\dfrac{\Omega_{r0}^{1/2}\xi_{min}}{\Gamma\,G\,\mu\,\xi}\right)^{3/2}
\label{eq:number.density}
\end{equation}
is the initial homogeneous density of loops and we have defined $\zeta \equiv \cos\alpha_i$ and $\lambda \equiv s_i/R_i$. This can be easily evaluated and is equal to
\begin{eqnarray}
N_{\text{tophat}}(\xi) &=& \dfrac{4\pi}{3}\,n(z_i,\xi) R_i^3\left[1 - \dfrac{3}{2}\left(\dfrac{\xi_{min}}{\xi}\right) + \dfrac{1}{2}\left(\dfrac{\xi_{min}}{\xi}\right)^3\right]\nonumber\\
&\approx& 1.0\,\mu_{-12}^{3/2}\left(\dfrac{1+z_v}{4}\right)^{15/2}\left(\dfrac{R_v}{60\,\text{kpc}}\right)^{9/2}\left(\dfrac{\xi_{min}}{\xi}\right)^{3/2}\left[1 - \dfrac{3}{2}\left(\dfrac{\xi_{min}}{\xi}\right) + \dfrac{1}{2}\left(\dfrac{\xi_{min}}{\xi}\right)^3\right]. \nonumber\\
\label{eq:Nwithin}
\end{eqnarray}
It rises very sharply and peaks at around $\xi \simeq 2.2\,\xi_{min}$, and agrees well with numerical simulations (see Fig. \ref{fig:number_loops_both}).

\section{Numerical simulations}
\label{sec:Simulation}

{Here} we shall first briefly discuss our simulation setup. We work with dimensionless quantities, so we define the positions and times in units of $R_v$ and $T_v$ respectively:
\begin{eqnarray}
\tilde{s} &\equiv& s/R_v\nonumber\\
\tau &\equiv& t/T_v.
\end{eqnarray}
In order to evolve loop trajectories, we need the background (time dependent) mass contained within the {radius of loop's current} location. Since there are no shell crossings in the top hat model (and hence mass within any comoving shell is conserved), we can invert Eq.\eqref{rtheta} to obtain the mass as a function of $\tilde{s}$ and $\tau$ (in units of $M_0$):
\begin{equation}
M(\tilde{s},\tau) = M_0\begin{cases}\left(\dfrac{\tilde{s}}{2\beta(\theta(\tau))}\right)^3 & \tilde{s} < R(t)/R_v\\ \left(\dfrac{\tilde{s}}{2\beta(\theta_2(\tau,\tilde{s}))}\right)^{3/4} & \tilde{s} > R(t)/R_v\end{cases}
\end{equation}
Here, $\theta(\tau)$ and $\theta_2(\tau,\tilde{s})$ are obtained by inverting
\begin{eqnarray}
\tau = \dfrac{d(\theta)}{D},\;\;\;\;\text{and}\;\;\;\;\tau\left(\dfrac{1}{\tilde{s}}\right)^{9/8} = \dfrac{d(\theta_2)}{D}\left(\dfrac{1}{2\beta(\theta_2)}\right)^{9/8}
\end{eqnarray}
respectively, and are monotonically increasing functions of their arguments (until the maximum value of $3\pi/2$ at virialization of the corresponding mass shell). With such a rescaling, we don't need to specify the initial overdensity $\delta_i \ll 1$ of the top hat, and the initial time $t_i$ (in the matter era) anymore.

As before, we neglect the initial loop velocity. This makes the problem effectively 2-dimensional and the motion of a loop can be restricted to $xy$-plane, with $x$ pointing in the rocket direction.  We can therefore use the coordinate definitions
\begin{eqnarray}
\tilde{x} &\equiv& \tilde{s}\cos\alpha,\nonumber\\
\tilde{y} &\equiv& \tilde{s}\sin\alpha,
\end{eqnarray}
where $\alpha$ is the angle between the loop's position vector and the rocket direction. Finally then, given the potential Eq.\eqref{potential}, loops obey the following equations of motion
\begin{eqnarray}
\tilde{x}'' &=& \dfrac{16\,D^2}{9}\left[-\dfrac{\tilde{x}}{\tilde{s}^3}\dfrac{M(\tilde{s},\tau)}{M_0} + \dfrac{\xi_*}{\xi}\right]\nonumber\\
\tilde{y}'' &=& \dfrac{16\,D^2}{9}\left[-\dfrac{\tilde{y}}{\tilde{s}^3}\dfrac{M(\tilde{s},\tau)}{M_0}\right],
\label{loopeom4}
\end{eqnarray}
where primes stand for derivatives with respect to $\tau$.

We scan various initial positions $\{{\tilde s}_i,\alpha_i\}$ and assign corresponding initial Hubble velocities in the radial directions, on top of the initial peculiar rocket velocity in the $x$-direction. For a given loop size and different initial conditions, we can be sure that once the virialization radius $\tilde{r}_{v}$ extends beyond the saddle point $\tilde{s}_{max}$ (which happens at times $\tau_{\text{max}} = (\xi/\xi_*)^{9/10}$), no loops that are outside of the bounding region, defined by the equipotential surface $V=E_B$, can ever be captured. The converse however is not true: some loops within the bounding region eventually escape.  We therefore extended the simulation beyond $\tau_{\text{max}}$ for sufficiently small values of $\xi$: we ran it until $\tau=10$ for $\xi\lesssim 14 \xi_* \sim 3\xi_{\text{min}}$ and until $\tau_{\text{max}}$ for larger loops. All loops of a given size, that remain within the bounding region at the end of the simulation are declared captured. All other loops are regarded as escaped. With the setup laid out, we now present our results.

\begin{figure}[t]
 \centering
\includegraphics[width=1\columnwidth]{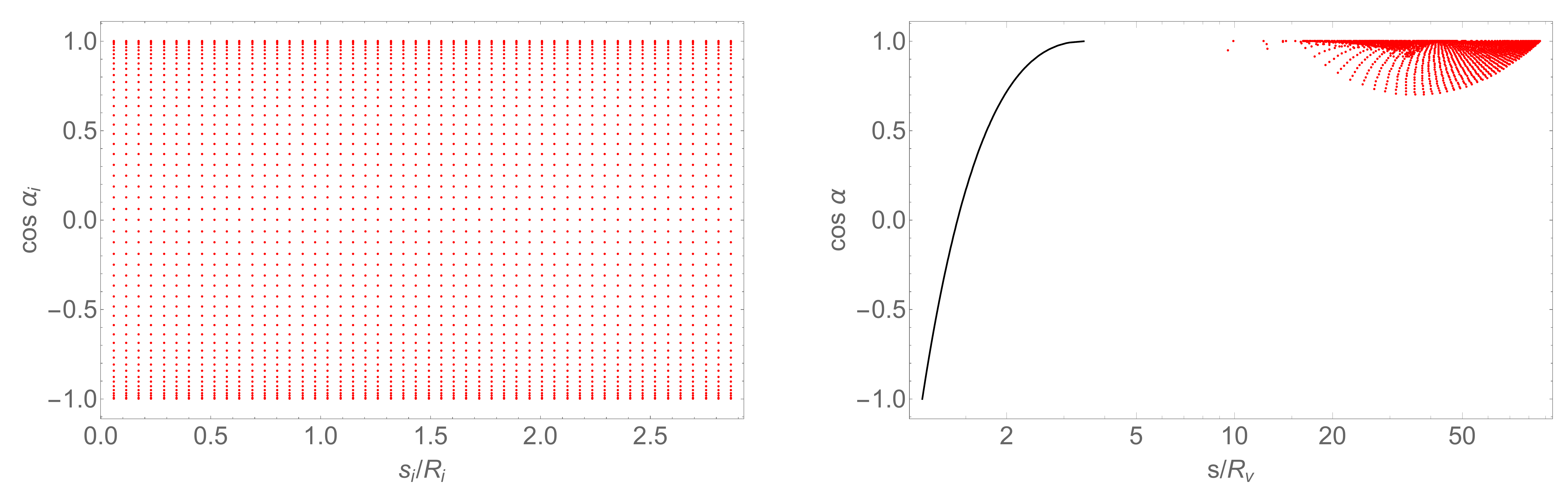}\vspace{0.1cm}\\
\caption{Loop distribution in the configuration space $\{s,\alpha\}$ for $\xi = 4.76\,\xi_{*}$ which is just a little smaller than $\xi_{min}$. Left panel shows the initial distribution, while the right panel shows all these loops at $t = 4\,T_v$. Black curve is the boundary of the bounding region. Red dots represent all the loops that escape the bounding region by the end of simulation. It is evident that no loops are captured.}
\label{fig:phase_space_initial_final_4_70_4Tv}
\end{figure}

\begin{figure}[t]
\centering
\includegraphics[width=1\columnwidth]{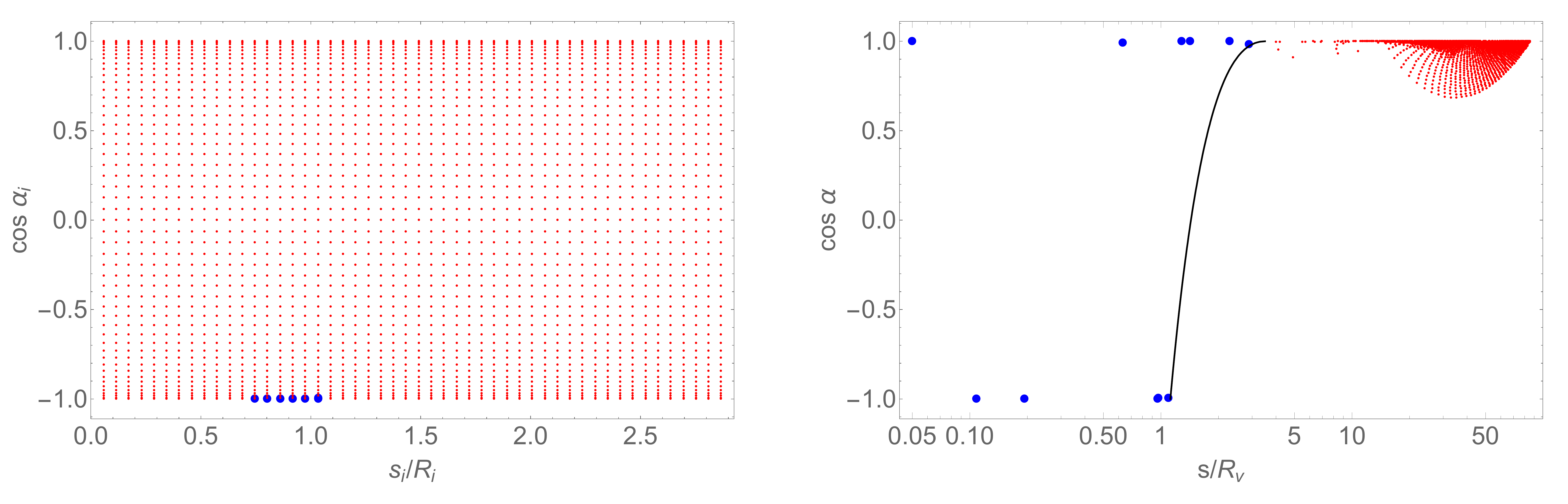}\vspace{0.1cm}\\
\caption{Same as in Fig.~\ref{fig:phase_space_initial_final_4_70_4Tv} for $\xi = 4.80\,\xi_{*}$ which is a little larger than $\xi_{min}$. Red and blue dots represent loops that escaped and that get captured, respectively. Only a few marginal loops are captured.}
\label{fig:phase_space_initial_final_4_80_4Tv}
\end{figure}

\begin{figure}[t]
\centering
\includegraphics[width=1\columnwidth]{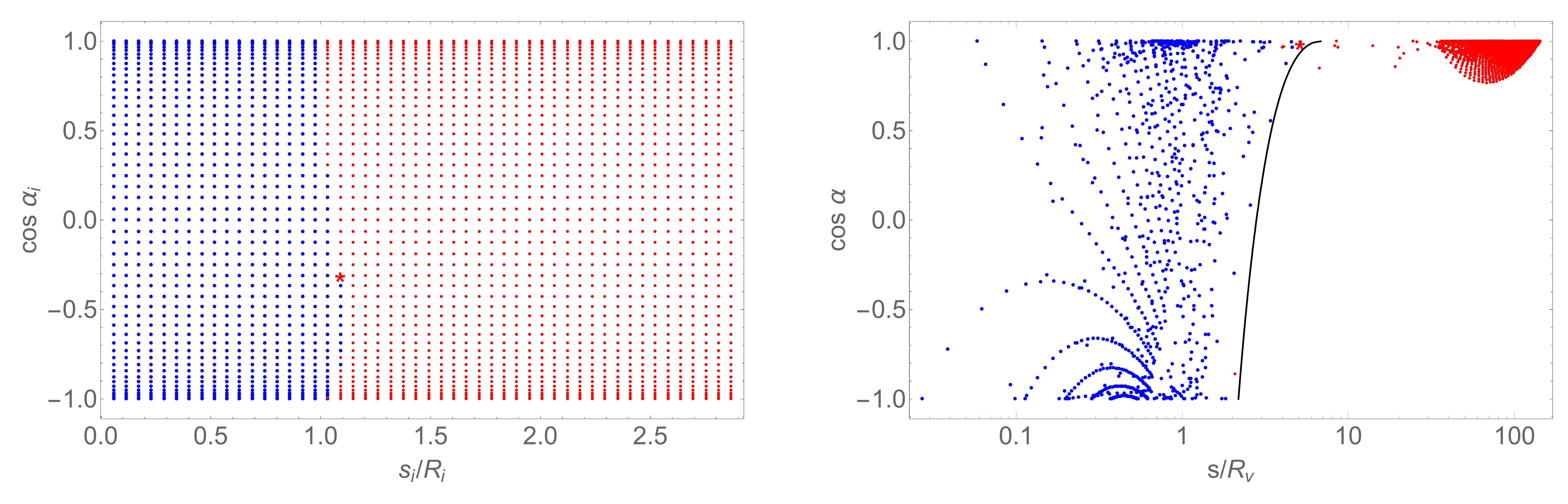}\vspace{0.1cm}\\
\caption{Same as in Fig.~\ref{fig:phase_space_initial_final_4_80_4Tv} for $\xi = 11.00\,\xi_{*} \approx 2.3 \xi_{min}$, except that the snapshot in the right panel is now at $t=8\,T_v$. Almost all loops that were initially within the co moving top hat radius get captured. The loop whose trajectory is shown in Fig.~\ref{fig:contour.plot}, is marked as a star.}
\label{fig:phase_space_initial_final_11_00_8Tv}
\end{figure}

We first present figures \ref{fig:phase_space_initial_final_4_70_4Tv}, \ref{fig:phase_space_initial_final_4_80_4Tv},\ref{fig:phase_space_initial_final_11_00_8Tv}, which illustrate loop dynamics for $\xi$ below and above $\xi_{min}$.  
We used $z_v = 3$ and $R_v = 60$~kpc, giving $\xi_{min}\approx 24.6$. The right panel of Fig.~\ref{fig:phase_space_initial_final_4_70_4Tv} shows the loop distribution at $t=4T_v$ resulting from the initial distribution shown in the left panel for $\xi \approx 24.5$, which is slightly below $\xi_{min}$.  Our color code is that blue and red dots represent loops that were captured and that escaped, respectively.  We see that no loops were captured in this case.    

Fig.~\ref{fig:phase_space_initial_final_4_80_4Tv} illustrates the same dynamics for $\xi \approx 24.7$, which is slightly above $\xi_{min}$. Here we see that almost all of the loops escape, with only a handful getting captured. Finally, Fig.~\ref{fig:phase_space_initial_final_11_00_8Tv} shows the loop distribution at $t=8 T_v$ with the same kind of initial setup and $\xi \approx 56.6$. In this case, most of the loops which were initially within the co moving top hat radius end up being captured. Note that a few loops within the bounding region at $t=8T_v$ are marked red, indicating that they escape by the end of simulation at $t=10T_v$.  This includes the loop whose trajectory is shown in Fig.~\ref{fig:contour.plot} and is highlighted with a red colored star. Such loops orbit the halo for a while before eventually escaping. These loops occupy a very small portion of the initial configuration space near the boundary of the top hat and thus have little effect on our results.

\begin{figure}[t]
\centering
\includegraphics[width=1\columnwidth]{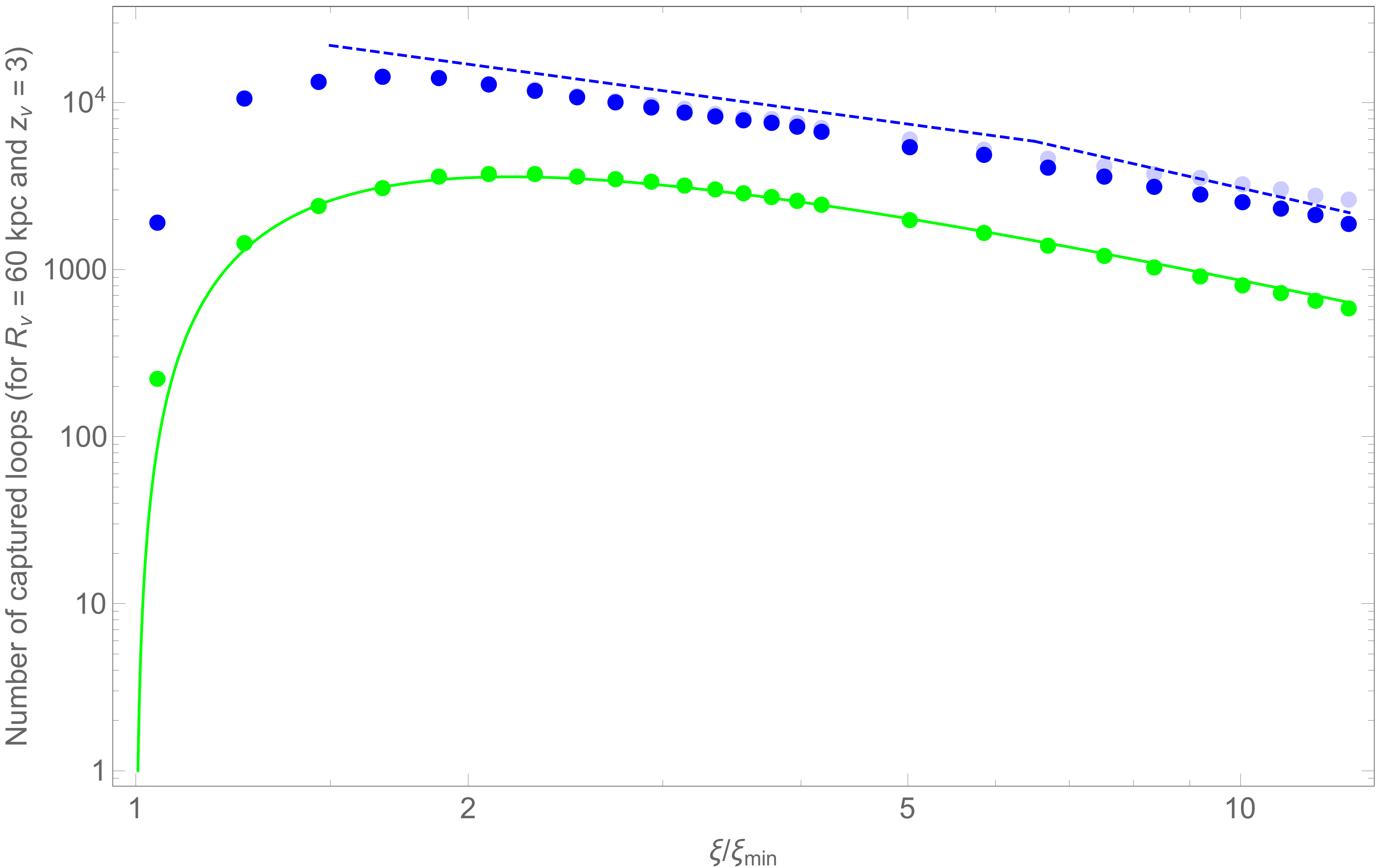}\vspace{0.1cm}\\
\caption{The number of loops captured as function of loop size $\xi$ for $G\mu = 10^{-15}$, $R_v = 60$ kpc, and $z_v = 3$. The lighter blue dots represent all captured loops assuming secondary infall continues until the end of the simulation. This gives $\sim \xi^{-9/10}$ behavior. The dark blue dots represent the subset of loops on which we have imposed the condition $r_i < r_{i\Lambda}$ with $r_{i\Lambda}$ from \eqref{ri_lambda} (that is, we discarded the loops which do not satisfy this condition). The green dots represent the subset of loops which remain within the top hat until $T_v$. The solid green line is the analytical estimate \eqref{eq:Nwithin}, while the dashed blue line is the analytical estimates \eqref{Ntot}, \eqref{NtotLambda}. For other values of $G\mu$, the data points and the analytic curves should be rescaled by a factor $(G\mu/10^{-15})^{-3/2}$.}
\label{fig:number_loops_both}
\end{figure}

Next, we plot the number of captured loops as a function of loop size $\xi$ in Figure \ref{fig:number_loops_both}.  The figure shows that the number of captured loops rises sharply as $\xi$ becomes bigger than $\xi_{min}$ and eventually dies out as $\xi^{-3/2}$. Therefore, the most abundant loops are of sizes $\sim \xi_{min}$. It is also apparent from this figure that our analytical estimates are quite accurate, both for the total number of captured loops, and for the subset that remained within the top hat until $T_v$ (of a given size).

The total number of captured loops (of any size) can be estimated as
\beq
N_{tot}\sim \int \frac{d\xi}{\xi} N(\xi).
\eeq
This can be integrated numerically using the dark blue data points in Fig.~\ref{fig:number_loops_both}, {and} the result is
\begin{eqnarray}
N_{tot} \sim 0.6\,\mu_{-12}^{-3/2}\left(\dfrac{R_v}{60\,\text{kpc}}\right)^{9/2}\left(\dfrac{1+z_v}{4}\right)^{15/2}.
\label{Ntot2}
\end{eqnarray} 

\section{Summary and discussion}
\label{sec:Conclusions}

We studied capture of cosmic string loops in collapsing dark matter halos using the spherical top hat model of halo formation.  We fully accounted for the rocket effect -- the loop acceleration due to asymmetric emission of gravitational waves by the loop -- and found that it does not prevent loop capture, provided that 
the string mass parameter $G\mu$ is sufficiently small and 
the loops are sufficiently large.  

We characterize the loop size by the dimensionless parameter $\xi=l/l_*\gtrsim 1$, where $l$ is the invariant length of the loop and $l_*\sim \Gamma G\mu t_0$ is the characteristic length of the smallest (and the most numerous) loops surviving at the present time $t_0$.  We find that loops can be captured {in the halo of a galaxy like the Milky Way} only if $\xi\gtrsim \xi_{min}\sim 25$. The unperturbed density of such loops is $\sim 100$ times smaller than that of the smallest loops with $\xi\sim 1$, and the expected number of captured loops is decreased correspondingly. The total number of loops captured in the halo 
is estimated as
\beq
N_{tot} \sim 0.6\,\left(\frac{G\mu}{10^{-12}}\right)^{-3/2}.
\label{NGmu}
\eeq

The dependence on $G\mu$ in Eq.~(\ref{NGmu}) is simply due to the fact that the initial unperturbed density of loops scales like $(G\mu)^{-3/2}$.
There are hardly any loops captured for $G\mu \sim 10^{-12}$, while a substantial number of them may get captured for smaller values. The most abundant size of captured loops within a halo is $\sim \xi_{min}$ given by \eqref{ximin}. Assuming that the loops are distributed more or less uniformly within the turnaround radius of the halo, $R_{ta}\sim 120~{\rm kpc}$, the average loop density within this radius is 
\beq
n_{ta}\sim 10^{-7} \left(\frac{G\mu}{10^{-12}}\right)^{-3/2} {\rm kpc}^{-3}. 
\eeq
It is about $10$ times higher than the present density of loops (of length $\xi\sim 1$) in the intergalactic space (but 100 times smaller than the density predicted by the Chernoff model at this distance from the Galactic center). This modest density enhancement can be understood as follows.  Most of the captured loops have sizes $\xi\sim 25$, so their density is reduced compared to that of the most numerous loops (with $\xi\sim 1$) by a factor $\xi^{-3/2}\sim 10^{-2}$. These loops are approximately comoving until the halo turnaround at $z_{ta}\approx 5$, so their density is enhanced by 
the factor (cf. Eq.~(\ref{tav})) $(9\pi^2/16)(1+z_{ta})^3\sim 10^3$.  Combining the two factors we obtain an order of magnitude enhancement.

We used a simple spherical model to describe the halo evolution. As we mentioned in Sec.~II.B, the choice of parameters $z_v$ and $R_v$ for this model is somewhat imprecise, since the predictions of the model do not provide accurate fits to observations or N-body simulations.  Even at the qualitative level, the model does not account for the hierarchical nature of structure formation.  According to hierarchical models, dark matter halos form by accretion and mergers of smaller halos.  It is possible then that the number of captured loops is larger than our estimates if smaller halos, formed at higher redshifts, capture loops more efficiently than the large halo of the galactic size. Furthermore, simulations suggest that dark matter halos are assembled from inside out, with dense central parts being assembled first \cite{Correa2015}. Then it is possible that the density of loops grows significantly towards the galactic center.  Chernoff \cite{Chernoff} suggests that it may grow proportionally to the dark matter density, in which case it would be 100 times higher at the location of the Sun ($r\sim 8~{\rm kpc}$) than at $r\sim 120~{\rm kpc}$. A definitive verdict on these issues would require combining numerical simulations of loop dynamics with N-body simulations of galaxy formation.  However, we believe that our analysis here can be used to yield some plausible answers.

\begin{figure}[t]
\centering
\includegraphics[width=1\columnwidth]{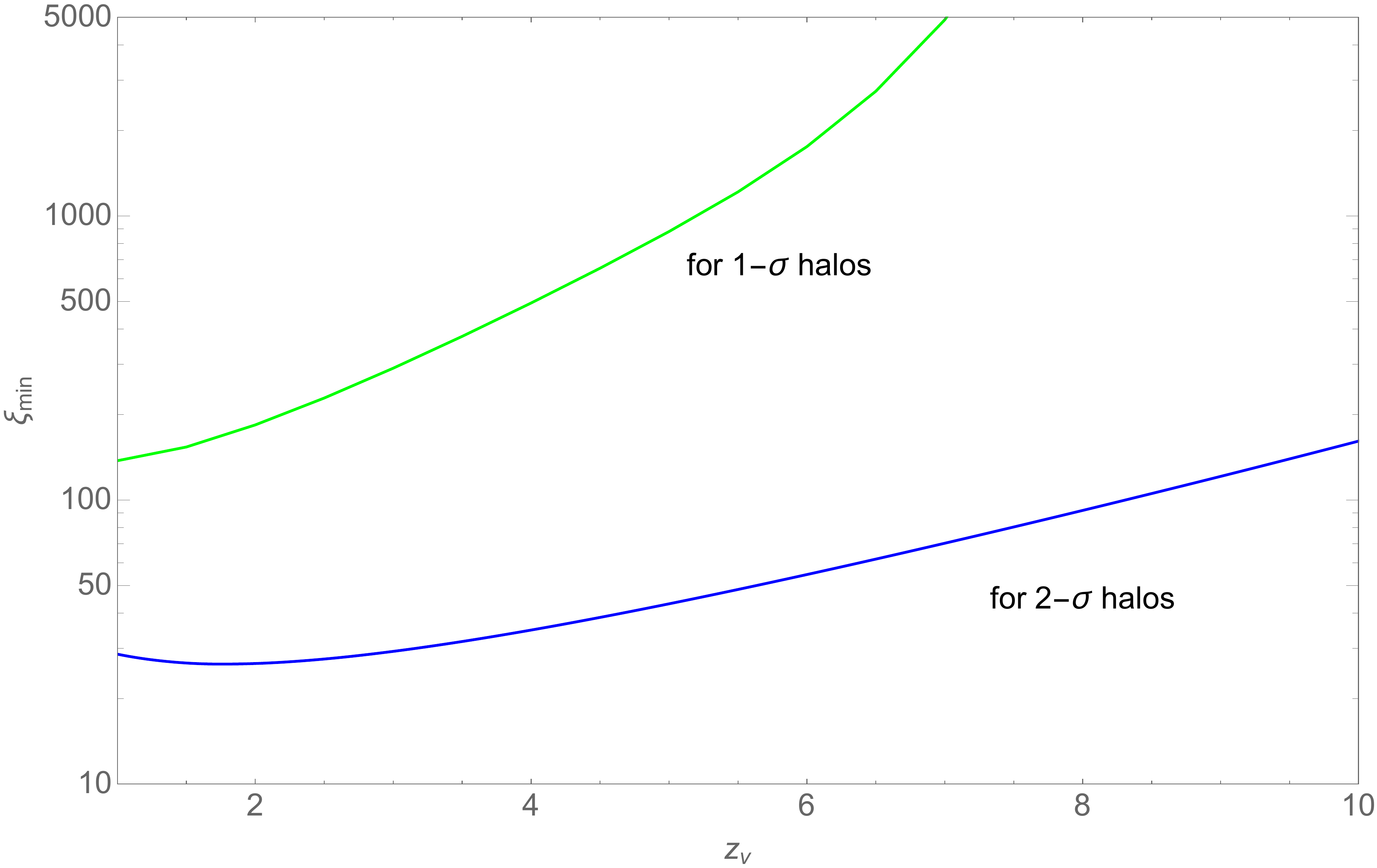}\vspace{0.1cm}\\
\caption{$\xi_{min}$ vs redshifts for $1\sigma$ and $2\sigma$ halos.}
\label{ximin12sigma}
\end{figure}

We first introduce the loop capture efficiency $\chi$, defined as the fraction of loops initially in the comoving halo which eventually get captured, with only loops surviving until present being counted. For halos virializing at redshift $z$, the efficiency is 
\beq
\chi(z)\sim \xi_{min}^{-3/2}(z)
\label{chi}
\eeq
where $\xi_{min}(z)=\xi_{min}(R_v(z),z)$ is given by Eq.~(\ref{ximin}) and $R_v(z)$ is the characteristic virialization radius of halos virializing at that redshift. The radius $R_v(z)$ can be estimated using the standard method of relating the top hat model to linear perturbation theory, as reviewed for example in Ref.~\cite{Loeb}. We plot the resulting quantity $\xi_{min}(z)$ in Fig.~\ref{ximin12sigma} for halos arising from $1\sigma$ and $2\sigma$ fluctuations.\footnote{For the calculation of $R_v(z)$ we used the cosmological parameters and the power spectrum of density fluctuations suggested by the best fit to the 9 year WMAP data, as given in \cite{Komatsu}.} In both cases it grows with the redshift, indicating that capture of loops in high-redshift halos is less efficient.  Furthermore, even though our galaxy might have originated from a $\sim 2\sigma$ fluctuation, the subsequent mergers are likely to be with typical, $1\sigma$ halos, for which $\xi_{min}(z)$ is further increased.  We conclude that loop capture is rather inefficient in the early halos, so most of the loops are captured during the later collapse of the galactic dark matter halo.  We therefore do not expect that accounting for  the hierarchical nature of galaxy formation would significantly modify our estimates of the total number of captured loops.

\begin{figure}[t]
\centering
\includegraphics[width=1\columnwidth]{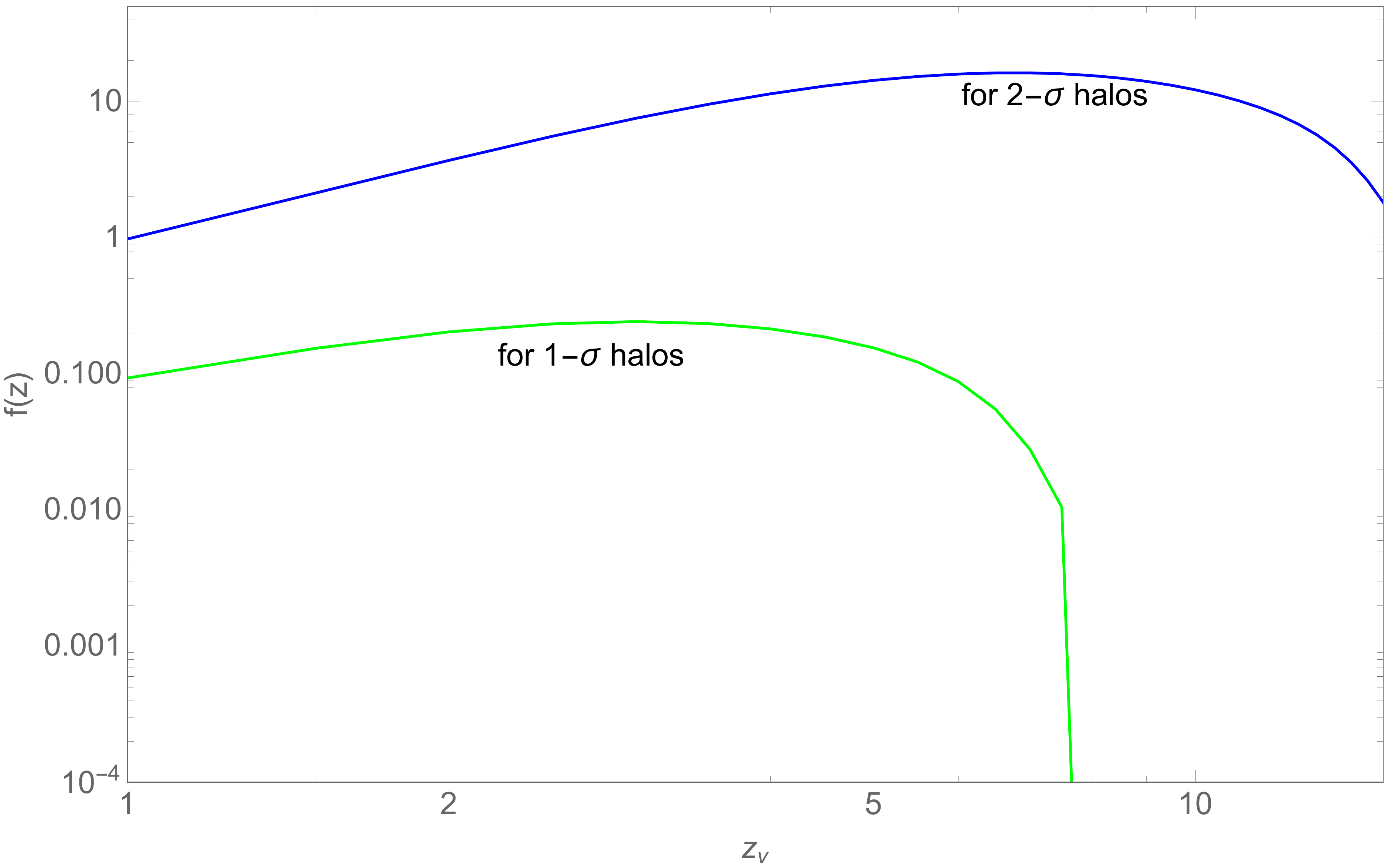}\vspace{0.1cm}\\
\caption{Density enhancement factor $f(z)$ of captured loops, for $1\sigma$ and $2\sigma$ halos.}
\label{fofz_new}
\end{figure}

A related but different issue is that of the loop density in early halos.  The unperturbed loop density for halos virializing at large $z$ is high ($\propto (1+z)^3$), and even if loops are captured at low efficiency, the loop density enhancement in early halos could be an increasing function of $z$.  This enhancement (compared to the average density of loops with $\xi\sim 1$ at present) is
\beq
f(z)\approx (9\pi^2/16) [1.5 (1+z)]^3 \xi_{min}^{-3/2}(z) ,
\eeq
where we have used that $(1+z_{ta})\approx 1.5 (1+z_v)$. It is plotted in Fig.~\ref{fofz_new} for $1\sigma$ and $2\sigma$ fluctuations. For $2\sigma$ halos we see that $f(z)$ grows by about a factor of 2 as the redshift varies from $z=3$ to $z\sim 5$, then it stays nearly flat until $z\sim 10$ and drops sharply at higher redshifts. For $1\sigma$ halos we have $f(z)<1$ in the entire range of $z$, so the density of captured loops in such halos is even smaller than the average density of loops in the intergalactic space. We conclude that the loop density in early halos is not substantially enhanced, and thus we do not expect a significant loop density enhancement towards the galactic center.

We now comment on some other simplifying assumptions that we adopted in our analysis.  

{\it (i)} We assumed that loops of a given length $l$ were formed at the same time $t_f\sim 10 l$ with initial velocities $v_f\sim 0.3$.  More realistically, loops are formed with a distribution of sizes and velocities. We do not expect this simplification to substantially affect our results.

\begin{figure}[t]
\centering
\includegraphics[width=1\columnwidth]{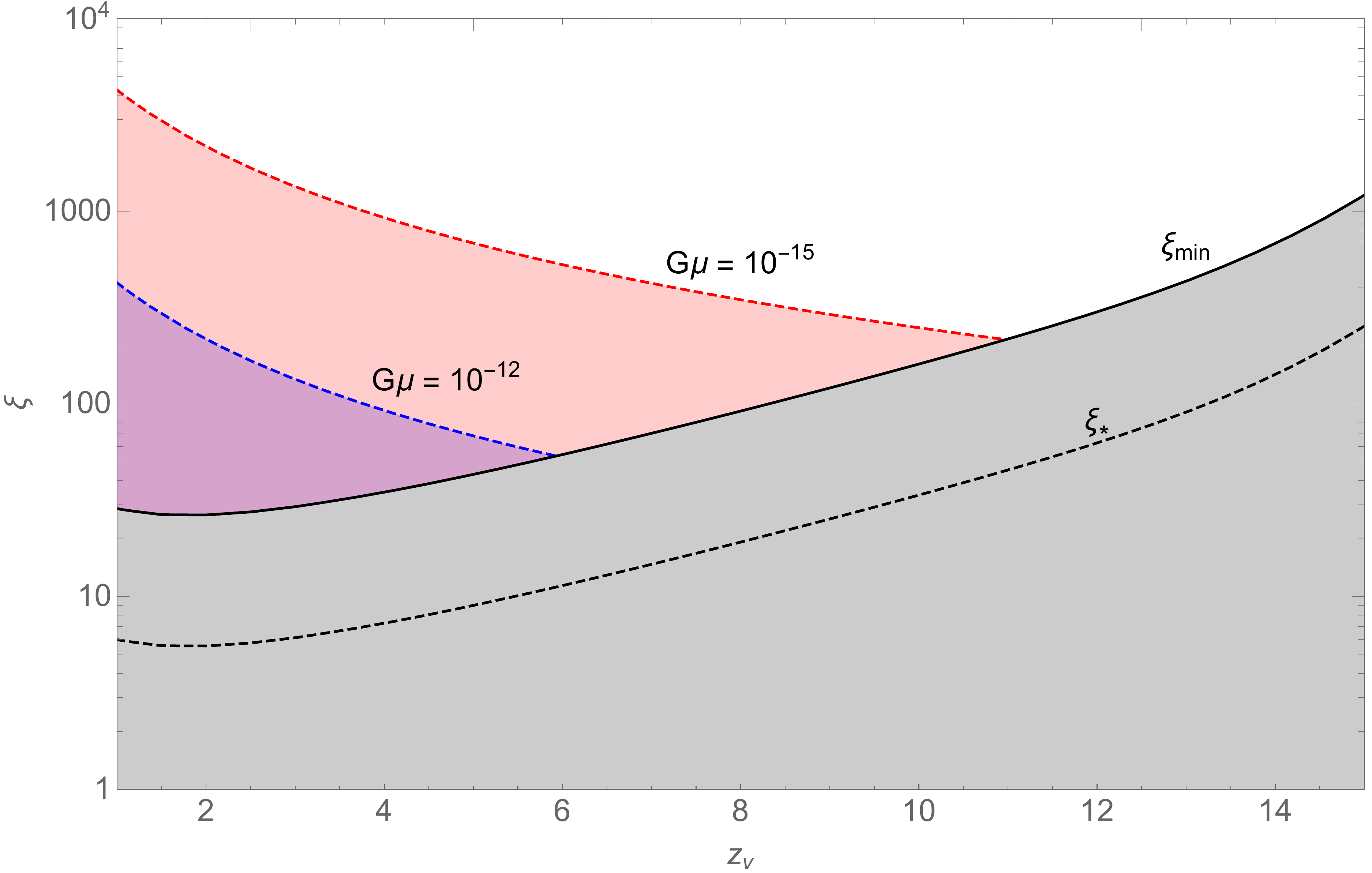}\vspace{0.1cm}\\
\caption{A plot summarizing various $\xi$'s. $\xi_{*}$ (and therefore also $\xi_{min}$) is obtained for $2\sigma$ halos using \cite{Komatsu}. Grey region is the `no capture' region where no loops can be captured. The blue and red dashed curves are the values of $\xi_{r}$ in Eq.\eqref{eq:xi_r} for $G\mu = 10^{-12}$ and $10^{-15}$ respectively, such that the containing shaded regions are where the rocket effect is dominant (and thus our estimates are accurate). Since most of the captured loops are of sizes $\sim \xi_{min}$, this plot shows that neglect of the initial peculiar velocities of loops is well justified.}
\label{variousxiplot}
\end{figure}

{\it (ii)} We assumed that the initial velocities are greatly redshifted by the time of halo collapse, so they can be neglected.  The condition for this to be justified is given by Eq.~(\ref{eq:xi_r}):
\beq
\xi \ll \xi_r \sim \mathcal{O}(10^2) \left(\frac{G\mu}{10^{-12}}\right)^{-1/3}\left(\frac{1+z_v}{4}\right)^{-5/3}.
\label{xir2}
\eeq
This condition is satisfied for galactic halos virializing at $z_v\lesssim 3$ and the most numerous captured loops (i.e. of sizes $\xi\sim \xi_{min}\sim 25$) for $G\mu\lesssim 10^{-12}$.  Note also that we find that almost no loops are captured for $G\mu\gtrsim 10^{-12}$ and that including loop's initial velocities can only decrease the number of captured loops.  We therefore expect our estimates for the number of captured loops to be accurate.  The range of values of $\xi$ and $z_v$ for which the condition (\ref{xir2}) is satisfied is shown in Fig.~\ref{variousxiplot} for representative values of $G\mu=10^{-12}$ and $10^{-15}$.

{\it (iii)} We assumed that the rates of energy and momentum radiation, characterized by the parameters $\Gamma$ and $\Gamma_p$, as well as the direction of the rocket force, remain constant throughout the relevant part of the loop's lifetime.  These parameters are expected to change on a timescale comparable to the lifetime, so this assumption is well justified, especially for large loops with $\xi\gg 1$.  We also assumed that the values of $\Gamma$ and $\Gamma_p$ are the same for all loops.  More realistically, we expect a distribution of values, and loops with smaller values of $\Gamma_p/\Gamma$ will have a smaller rocket force and will be more readily captured.  The distribution for $\Gamma_p/\Gamma$ is presently unknown; it may significantly influence the loop capture.

We finally summarize the differences of our results from those of Chernoff \cite{Chernoff}.  Chernoff found that clustering of loops is essentially independent of their size and that the density distribution of loops in the galaxy follows that of dark matter, with an overall correction factor $\beta$ which depends only on $G\mu$.  As $G\mu$ varies from $\sim 10^{-10}$ to $\sim 10^{-15}$, $\beta$ changes from near zero to $0.4$ and saturates at that value for $G\mu<10^{-15}$. This picture is rather different from our conclusions. The main reason for this discrepancy is that Chernoff ignores the rocket effect before and during halo collapse and only considers its role for ejection of captured loops.  On the other hand, our analysis shows that the rocket effect plays a dominant role for $G\mu\lesssim 10^{-12}$, so we adopted the opposite approximation of neglecting loop initial velocities.

These differences have important implications for observational effects of loop clustering.  According to our picture, the distance to the nearest loop is
\beq
d\sim n_{ta}^{-1/3} \sim 200 \left(\frac{G\mu}{10^{-12}}\right)^{1/2} {\rm kpc},
\eeq
while Chernoff's picture gives $d\sim 10(G\mu/10^{-12})~$kpc at the Sun's location. With our estimates, detection of nearby loops by their gravitational wave signal or by microlensing of stars would be more challenging than suggested by Refs.~\cite{ChernoffTye,ChernoffTye2,Chernoff3,Hogan,Hogan2}.

It should be noted that loop density could be significantly enhanced if instead of "ordinary" field theory strings (which we assume here) one considers cosmic superstrings \cite{ChernoffTye,ChernoffTye2}.  This is due to the following two factors: superstrings have a low reconnection probability, resulting in a higher density of loops, and superstring models typically predict the formation of a number of different string species.  We also note that the observational implications of loop clustering have been discussed so far only in relation to gravitational effects of strings.  But cosmic strings are likely to be superconducting \cite{Witten}, in which case they can have a nontrivial interaction with the magnetic field of the Galaxy.  The resulting observational effects may be more easily detectable.  This issue deserves further investigation.

\section{Acknowledgements}

{We are grateful to Jose Blanco-Pillado, Andrei Gruzinov, Avi Loeb, Ken Olum and Levon Pogosian for very useful discusions and comments on the initial draft of the paper.  
This work was supported in part by the National Science Foundation under grant PHY-1820872.}


\begin{thebibliography}{99}

\bibitem{Book}
A. Vilenkin and E. P. S. Shellard, {\it Cosmic Strings and Other Topological Defects}
(Cambridge University Press, Cambridge, 2000).

\bibitem{Tanmay}
T. Vachaspati, L. Pogosian and Daniele Steer, Scholarpedia, 10 (2) : 31682 (2015).

\bibitem{ChernoffTye}
D.~F.~Chernoff and S.~H.~Tye, Int.~J.~Mod.~Phys., D24, 1530010 (2015).
   
\bibitem{Chernoff}
D.~F.~Chernoff, "Clustering of superstring loops," arXiv:0908.4077 [astro-ph].   
   
\bibitem{ChernoffTye2}
D.~F.~Chernoff and S.~H.~Tye,~"Detection of low tension cosmic superstrings," JCAP 1805, 002 (2018).

\bibitem{Chernoff3}
D.~F.~Chernoff, A.~Goobar and J.~J.~Renk, "Prospects of cosmic superstring detection through microlensing of extragalactic point-like sources", Mon.~Not.~Roy.~Astron.~Soc. {\bf 491}, 596 (2020).

\bibitem{Hogan}
M.~R.~DePies and C.~J.~Hogan, "Harmonic gravitational waves spectra of cosmic string loops in the galaxy," arXiv:0904.1052 [astro-ph].

\bibitem{Hogan2}
Z.~Khakhaleva-Li and C.~J.~Hogan, "Will LISA detect harmonic gravitational waves from galactic cosmic string loops?", arXiv:2006.00438 [astro-ph.CO]. 

\bibitem{GunnGott}
J.~E.~Gunn and J.~R.~Gott, "On the infall of matter into clusters of galaxies and some effects on their evolution," Astrophys.~J.~{\bf 176}, 1 (1972).

\bibitem{Bertschinger}
E.~Bertschinger, "Self-similar secondary infall and accretion in an Einstein-de Sitter universe," ~Astrophys.~J.~Suppl.~Series,~{\bf 58}, 39 (1985).

\bibitem{NFW}
J.~F.~Navarro, C.~S.~Frenk and S.~D.~White, "The structure of cold dark matter halos," Astrophys.~J.~{\bf 462}, 563 (1996).

\bibitem{Suto}
D.~Suto {\it et. al.}, "Confrontation of top-hat spherical collapse against dark halos from cosmological N-body simulations,"
Publ.~Astron.~Soc.~Jap.~ 68, 14 (2016)~.

\bibitem{Evans}
T.~A.~Evans {\it et al}, ``How unusual is the Milky Way's assembly history?," 
[arXiv:2005.04969 [astro-ph.GA]].

\bibitem{Loeb}
  R.~Barkana and A.~Loeb,
  ``In the beginning: The First sources of light and the reionization of the Universe,''
  Phys.\ Rept.\  {\bf 349}, 125 (2001)
    [astro-ph/0010468].

\bibitem{Komatsu}
E.~Komatsu, https://wwwmpa.mpa-garching.mpg.de/~komatsu/crl/list-of-routines.html

\bibitem{BOS16}
J. J. Blanco-Pillado, K. D. Olum and B. Shlaer, "The number of cosmic string loops," Phys. Rev. D {\bf 89}, 023512 (2014).

\bibitem{BOS18}
J.~J.~ Blanco-Pillado, K.~D.~Olum, X.~Siemens, "New limits on cosmic strings from gravitational wave observation,"
Phys.~Lett.~B 778, 392 (2018).

\bibitem{BO17}
J.~J.~ Blanco-Pillado and K.~D.~Olum, "Stochastic gravitational wave background from smoothed cosmic string loops,"
Phys.~Rev.~D {\bf 96}, 104046 (2017)
[arXiv:1709.02693[astro-ph-CO]].

\bibitem{HoganRees}
C. ~J.~Hogan and M.~J.~Rees,  "Gravitational interactions of cosmic strings,"~Nature {\bf 311}, 109 (1984).

\bibitem{VV}
T.~Vachaspati and A.~Vilenkin, "Gravitational radiation from cosmic strings," Phys.~Rev. D {\bf 31}, 3052 (1985).


\bibitem{Correa2015}
C.~A.~Correa {\it et al}, "The accretion history of dark matter halos - II. The connection with the mass power spectrum and the density profile," Mon.~Not.~Roy.~Astron.~Soc. 450, 1521 (2015)
[arXiv:1501.04382 [astro-ph.CO]]

\bibitem{Witten}
E.~Witten, "Superconducting strings," Nucl. Phys. B 249, 557 (1985).

\end{thebibliography}
\end{document}